\documentclass[runningheads]{llncs}

\usepackage[T1]{fontenc}
\usepackage[numbers]{natbib}
\usepackage{multirow}
\usepackage{graphicx}
\usepackage{tabularx}    % for auto-wrapping to \linewidth
\usepackage{array}       % for \newcolumntype and 
\providecommand{\Description}[1]{}
\usepackage{float} % for [H]
\usepackage{threeparttable} % for tablenotes environment
\usepackage[table]{xcolor} % needed for \rowcolor
\usepackage{float} \usepackage{hyperref}
\usepackage{orcidlink}

\begin{document}

%\title{Hope, Aspirations, and the Impact of LLMs on Women Programmers in Afghanistan}
\title{Hope, Aspirations, and the Impact of LLMs on Female Programming Learners in Afghanistan\thanks{Preprint. This paper has been accepted to the Workshop on Artificial Intelligence with and for Learning Science 2025 (WAILS-2025); please cite that version instead.}}

%\titlerunning{Hope, Aspirations, and the Impact of LLMs on Women Programmers in Afghanistan}
\titlerunning{Hope, Aspirations, and the Impact of LLMs on Female Programming Learners in Afghanistan}

% \author{Hamayoon Behmanush\inst{1}\and
% Freshta Akhtari\inst{2} \and
% Roghieh Nooripour\inst{3} \and
% Ingmar Weber\inst{1} \and
% Vikram Kamath Cannanure\inst{1}}

\author{Hamayoon Behmanush\inst{1}\orcidlink{0000-0003-0995-7207} \and
Freshta Akhtari\inst{2} \orcidlink{0009-0009-1722-977X} \and
Roghieh Nooripour\inst{3} \orcidlink{0000-0002-5677-0894} \and
Ingmar Weber\inst{1} \orcidlink{0000-0003-4169-2579} \and
Vikram Kamath Cannanure\inst{1} \orcidlink{0000-0002-0944-7074}}

\authorrunning{Behmanush et al.}

\institute{Saarland Informatics Campus, Saarland University, Saarbrücken, Germany \\  
\email{\{behmanush, iweber, cannanure\}@cs.uni-saarland.de} \and
Computer Science Faculty, Parwan University, Charikar, Afghanistan \and
Department of Counseling, Qazvin Branch, Islamic Azad University, Qazvin, Iran}

\maketitle
\begin{abstract}
Designing impactful educational technologies in contexts of socio-political instability requires a nuanced understanding of educational aspirations. Currently, scalable metrics for measuring aspirations are limited. This study adapts, translates, and evaluates Snyder’s Hope Scale \cite{snyder1991will} as a metric for measuring aspirations among 136 women learning programming online during a period of systemic educational restrictions in Afghanistan. The adapted scale demonstrated good reliability (Cronbach’s $\alpha = 0.78$) and participants rated it as understandable and relevant. While overall aspiration-related scores did not differ significantly by access to Large Language Models (LLMs), those with access reported marginally higher scores on the \textit{Avenues} subscale ($p = .056$), suggesting broader perceived pathways to achieving educational aspirations. These findings support the use of the adapted scale as a metric for aspirations in contexts of socio-political instability. More broadly, the adapted scale can be used to evaluate the impact of aspiration-driven design of educational technologies.

\end{abstract}

\section{Introduction}
Understanding learners’ contexts is essential for designing educational technologies. Scholars in Human–Computer Interaction for Development (HCI4D) and educational technology advocate for an aspirations-based approach to technology design \cite{toyama2018needs}, emphasizing that aligning technology with learners’ long-term goals can foster sustainable socio-economic impact. Aspirations are defined as goals that extend one’s current circumstances \cite{toyama2018needs} and are described through two subscales: \textit{Agency} (the individual’s capacity and determination to pursue long-term goals) and \textit{Avenues} (the perceived opportunities and pathways available within socio-structural conditions) \cite{toyama2018needs, kumar2014facebook}. Meanwhile, socio-political challenges play an influential role in shaping aspirations, as they disrupt social order through political change or restrictive norms \cite{laferrara2019aspirations}. In the context of educational technology, prior work has attempted to measure aspirations using the theory of planned behavior for avenues and agency with little success \cite{cannanure2023dia}. In addition, conceptualizing hope as comprising aspirations, agency, and pathways \cite{lybbert2016hope} may provide a clearer mapping for measuring aspiration. Building on this, the Hope Scale \cite{snyder1991will}, a user-administered instrument with two dimensions — Agency (goal-directed determination) and Pathways (perceived capacity to generate routes to goals) — may serve as an aspiration-aligned metric of progress toward learners’ aspirations. However, despite its relevance, the Hope Scale remains underutilized in technological interventions as a metric for aspirations, particularly in settings where socio-political constraints shape learners’ aspirations.

Emerging technologies, such as large language models (LLMs), are increasingly being utilized in education. Prior studies show that instability can hinder education by increasing psychological stress and exacerbating the digital divide \cite{reich2020failure}. At the same time, LLMs offer promising educational support, including personalized tutoring \cite{kasneci2023chatgpt}, instant feedback \cite{dai2022educational}, and reinforcement of self-efficacy \cite{wang2023adaptive}. Although research has examined how technology can foster educational resilience \cite{behmanush2025genai} and act as an amplifier of socio-economic change \cite{toyama2011technology} in unstable contexts, the role of LLMs in shaping learners’ aspirations, particularly among marginalized populations, remains underexplored.

To address these gaps, this study investigates two research questions:

\begin{description}
\item[RQ1:] What is the feasibility of using the Snyder Hope Scale to measure the educational aspirations of adult women in contexts of socio-political instability?
\item[RQ2:] Does access to Large Language Models (LLMs) influence the educational aspirations of these learners?
\end{description}

To answer these questions, we surveyed 136 women in Afghanistan who were studying programming online during periods of educational and employment restrictions. We adapted the Hope Scale to the programming context, translated it into Persian for accessibility, and applied statistical methods such as Cronbach’s alpha \cite{boateng2018best} and ANOVA \cite{pandey2021measurement} to assess both the reliability of the adapted scale and the influence of LLM access on aspirations.

Our study makes two primary contributions: (1) we evaluate the feasibility and reliability of an adapted Hope Scale in a marginalized and unstable context; and (2) we assess the relationship between LLM access and learner aspirations. We further reflect on aspiration-driven design and methods for impact assessment in educational technology.

\section{Related Work}

\subsection{Technology, Aspirations, and Measurement Challenges}
Research in educational technology and HCI4D demonstrates that digital tools can shape learners’ long-term goals by expanding access to information, education, and skills support \cite{behmanush2025genai, donner2015after}. Consistent with an aspirations-based approach to technology design, which emphasizes aligning technology with learners’ long-term goals to foster sustainable socio-economic impact \cite{toyama2018needs}, aspirations can be understood as goals that extend one’s current circumstances and are structured by two subscales: \textit{Agency} (capacity and determination to pursue long-term goals) and \textit{Avenues} (perceived opportunities within socio-structural conditions) \cite{toyama2018needs, kumar2014facebook}. Socio-political instability further shapes these aspirations by disrupting social order through political change or restrictive norms \cite{laferrara2019aspirations}.

A persistent challenge is measuring aspirations in educational settings amid socio-political disruptions. Prior work in education and development studies often relied on proxies such as occupational goals or long-term study plans \cite{schoon2002teenage}. In HCI and related fields, aspirations have been modeled using the Aspirations–Avenues–Agency framework \cite{kumar2014facebook} and adaptations of the Theory of Planned Behavior, with limited success in capturing avenues and agency in practice \cite{cannanure2023dia}. Building on the view that hope comprises aspirations, agency, and pathways \cite{lybbert2016hope}, we adopt Snyder’s Hope Scale \cite{snyder1991will}, a user-administered instrument with two dimensions, \textit{Agency} (goal-directed determination) and \textit{Pathways} (capacity to generate routes to goals), as an aspiration-aligned metric for measuring learners’ aspirations. The Hope Scale is widely used to assess agency and pathways \cite{boateng2018best, hansen2020measuring} yet remains underutilized in technological interventions as a measure of aspiration, particularly where socio-political constraints shape learners’ goals. In our mapping, Hope-\textit{Agency} corresponds to the aspirations subscale of \textit{Agency}, and Hope-\textit{Pathways} corresponds to \textit{Avenues} \cite{toyama2018needs, lybbert2016hope}, providing a basis for evaluating how aspiration can be measured and educational technologies support learners’ long-term trajectories.

%yet it remains unclear whether the Hope Scale adequately captures educational aspirations in socio-politically unstable settings or whether it can be used to evaluate the effects of technological interventions on learners’ long-term trajectories.

\subsection{Large Language Models and Educational Aspirations}
Large Language Models (LLMs), such as ChatGPT \cite{chatgpt}, are rapidly transforming educational practice. Their integration has extended across domains, supporting personalized tutoring and adaptive learning \cite{kasneci2023chatgpt}, writing assistance and automated feedback \cite{dai2022educational}, and research support through information access \cite{rudolph2023chatgpt}. Beyond academic skills, studies suggest that LLM-based tools can strengthen self-efficacy \cite{wang2023adaptive}, increase intrinsic motivation, and broaden access to high-quality educational materials \cite{chan2023ai}. Together, these affordances position LLMs as potentially influential not only in shaping immediate learning outcomes but also in supporting longer-term educational trajectories.  

These possibilities are particularly salient in contexts where formal education is disrupted by socio-political instability \cite{behmanush2025genai}. Prior work in HCI4D has shown how technology can buffer such disruptions: open educational resources and mobile platforms have enabled learners in under-resourced communities to sustain engagement through flexible and affordable access \cite{ally2013open}. Recent studies suggest that LLMs may extend this resilience-building role, offering opportunities for continued learning even when institutional support is weakened \cite{behmanush2025genai, lecce2024reflections}. Yet despite increasing evidence of their pedagogical benefits, little is known about how LLMs influence learners’ aspirations, particularly for marginalized populations whose educational and career goals are most vulnerable to socio-political crises.

\section{Methodology}
We surveyed 136 women in Afghanistan whose educational and career aspirations are constrained by ongoing political restrictions. To ensure contextual fit,  the Hope Scale \cite{snyder1991will} was revised to better align with programming education and to account for the lived experiences of learners in Afghanistan. The study received approval from the Ethical Review Board (ERB) of the Faculty of Mathematics and Computer Science at Saarland University (No. 23-10-7), and all participants provided informed consent.

\subsection{Participants}
Our study included 136 women who were studying programming online during a period of formal educational and employment restrictions. Participants ranged in age from 18 to 40 years (M = 22.8, SD = 3.41), reflecting a predominantly young female cohort. Table \ref{tab:my-table} provides a comprehensive demographic overview of the survey participants.
\begin{table}[H]
\centering
\setlength{\tabcolsep}{12pt}
\renewcommand{\arraystretch}{1.2}

\caption{Demographics of study participants.}
\label{tab:my-table}

\begin{tabular}{l l}
\hline
\textbf{Question} & \textbf{Response (Count, \%)} \\
\hline
\multirow{3}{*}{Age Group} 
    & 18--21 \hfill 47 (35\%) \\
    & 22--25 \hfill 65 (48\%) \\
    & 26+ \hfill 24 (17\%) \\
\hline
\multirow{3}{*}{Education Completed} 
    & Bachelor's Degree \hfill 63 (46\%) \\
    & High School  \hfill 58 (43\%) \\
    & Other \hfill 15 (11\%) \\
\hline
\multirow{4}{*}{Internet Connection} 
    & Mobile Data \hfill 85 (62\%) \\
    & Satellite \hfill 20 (15\%) \\
    & DSL \hfill 19 (14\%) \\
    & Other \hfill 12 (9\%) \\
\hline
\end{tabular}
\end{table}

\subsection{Data Collection}
Following the ERB approval, we adapted the Hope Scale \cite{snyder1991will} to reflect the programming context. The adaptation was reviewed by the third author, who is a specialist in the use of such scales in research contexts. Both the original and adapted items are listed in Table \ref{tab:hope-scale}. To ensure accessibility, the items were translated into Persian, and the scale was piloted with 10 participants before the main survey.  

Participants were recruited through three NGOs that support women in online programming education, enabling safe access to a population that is otherwise hard to reach. The adapted survey was distributed via Google Forms to 225 potential participants, yielding 136 complete responses (a response rate of $\approx$60\%). To evaluate the understandability and relevance of the adapted items to the programming context, we conducted a follow-up post-survey with 20 randomly selected participants.

\subsection{Data Analysis}
We assessed the reliability of the adapted scale using Cronbach’s alpha \cite{boateng2018best}. Following Hansen et al. \cite{hansen2020measuring}, responses were scored on a 5-point Likert scale, producing overall aspiration scores as well as subscale scores for \textit{Agency} and \textit{Avenues} (corresponding to the Hope Scale’s “agency” and “pathways” components). To explore differences across groups, we applied one-way ANOVA \cite{pandey2021measurement}, examining whether factors such as education level, age, or access to LLMs influenced aspiration scores.

\section{Results}
\subsection{The Adapted Hope Scale as a Feasible Metric for Measuring Educational Aspirations}
The adapted Hope Scale consisted of eight core items, four measuring \textit{Agency} and four measuring \textit{Avenues} (corresponding to the “pathways” component in the original scale), as well as four filler items excluded from analysis (see Table \ref{tab:hope-scale}). To improve clarity for the target population, responses were collected using a 5-point Likert scale, rather than the 6-point version used by \cite{hansen2020measuring} or the original 8-point version. Total aspiration scores ranged from 8 (lowest) to 40 (highest), with subscale scores ranging from 4 to 20.

\begin{figure}[]
  \centering
  \includegraphics[width=0.7\linewidth]{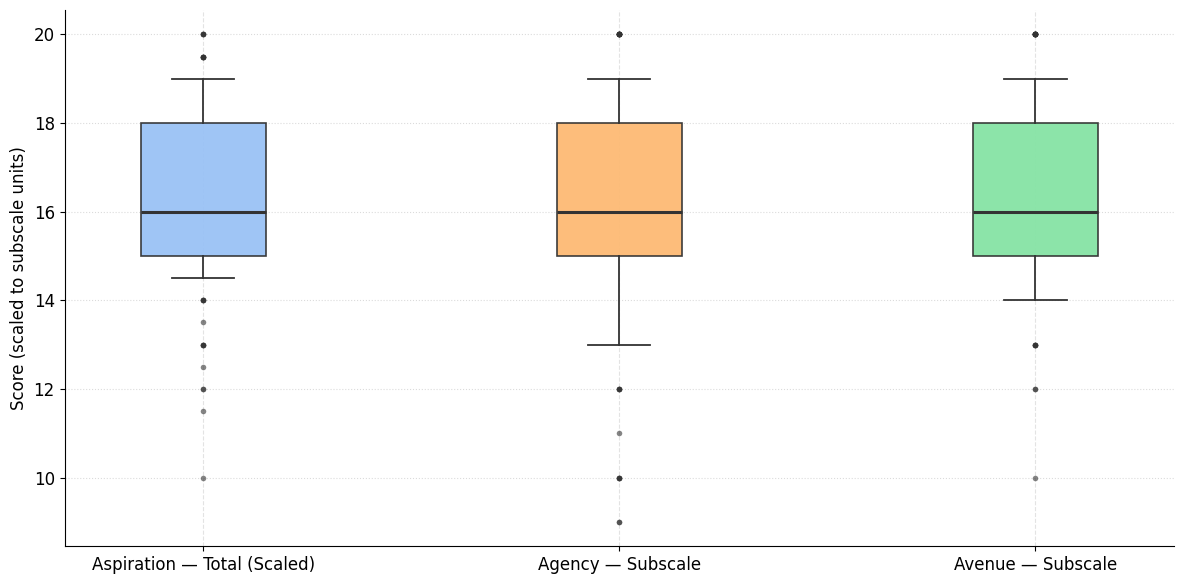}
  \caption{The box plots summarize score distributions for Aspiration and its subscales (\textit{Agency} and \textit{Avenues}). Each box represents the interquartile range (25th–75th percentiles) with the median marked by the line inside. Whiskers extend to the 10th and 90th percentiles.}
  \label{ScoreDistribution}
\end{figure}

To assess internal consistency, we calculated Cronbach’s alpha \cite{boateng2018best}. The overall scale demonstrated good reliability ($\alpha = 0.78$). The \textit{Avenues} subscale, reflecting respondents’ perceived ability to identify multiple routes toward achieving goals, yielded an alpha of 0.68. The \textit{Agency} subscale, capturing determination and self-belief in pursuing goals, scored 0.67. Although these are slightly below the conventional 0.70 threshold, methodological reviews suggest they may still be acceptable in exploratory research and in challenging field contexts \cite{Taber2018}. 

Figure \ref{ScoreDistribution} illustrates the distribution of aspiration and subscale scores. Distributions were moderately skewed toward higher values, suggesting that many participants reported elevated aspiration levels. Table \ref{tab:desc_stats} supports these observations by providing descriptive statistics: the mean aspiration score was 32.7 ($SD = 3.9$), with mean subscale scores of 16.1 ($SD = 2.5$) for \textit{Agency} and 16.6 ($SD = 2.1$) for \textit{Avenues}. Results from the post-survey, administered to a random sample of 20 participants, indicate that the adapted questions were largely understandable and relevant. Participants rated each item individually, and 91.2\% reported fully understanding the questions, while 72\% rated them relevant to their field of study.

\begin{table}[H]
\centering
\setlength{\tabcolsep}{10pt} % increase column separation
\renewcommand{\arraystretch}{1.2} % increase row spacing a bit

\caption{Descriptive statistics of Aspiration and subscale scores.}
\label{tab:desc_stats}

\begin{tabular}{l r r r}
\hline
\textbf{Statistic} & \textbf{Aspiration (Total)} & \textbf{Agency} & \textbf{Avenues} \\
\hline
Count   & 136   & 136   & 136 \\
Mean    & 32.7 & 16.1 & 16.6 \\
SD      & 3.9  & 2.5  & 2.1 \\
Min     & 20    & 9     & 10 \\
25\%    & 30    & 15    & 15 \\
Median  & 32    & 16    & 16 \\
75\%    & 36    & 18    & 18 \\
Max     & 40    & 20    & 20 \\
\hline
\end{tabular}
\end{table}

\subsection{Marginal Impact of LLMs on Educational Aspirations}
\label{sec:finding2}
We investigated whether access to Large Language Models (LLMs) and demographic factors were associated with differences in aspiration-related scores. Demographic variables such as age and education level showed no significant effects ($p = .593$ and $p = .378$, respectively). A one-way ANOVA was conducted to compare scores for the \textit{Agency} subscale, \textit{Avenues} subscale, and overall Aspiration between participants with and without LLM access. As shown in Table \ref{tab:desc_stats1}, no statistically significant differences were observed. Agency scores did not differ between groups ($p = .423$), nor did overall aspiration scores ($p = .129$). However, the Avenues subscale approached marginal significance ($p = .056$), suggesting a possible trend toward broader perceptions of pathways among participants with LLM access. Although this suggests a positive effect of LLM access on aspiration, the results should be interpreted with caution, as the overall aspiration scores did not reach statistical significance.

\begin{table}[h!]
\centering
\setlength{\tabcolsep}{10pt} % increase column separation
\renewcommand{\arraystretch}{1.2} % increase row spacing a bit

\caption{One-way ANOVA results for Aspiration-related scores by LLM access group.}
\label{tab:desc_stats1}

\begin{tabular}{l l l l}
\hline
\textbf{Scale} & \textbf{Group} & \textbf{Mean (SD)} & \textbf{N} \\
\hline
\multirow{2}{*}{Agency} 
  & Don’t have access & 15.9 (2.6) & 63 \\
  & Have access       & 16.3 (2.4) & 73 \\
  & \multicolumn{3}{r}{\textit{F (p) = 0.65 (0.423)}} \\
\hline
\multirow{2}{*}{Avenues} 
  & Don’t have access & 16.3 (2.0) & 63 \\
  & Have access       & 16.9 (2.1) & 73 \\
  & \multicolumn{3}{r}{\textit{F (p) = 3.70 (\textbf{0.056})}} \\
\hline
\multirow{2}{*}{Aspiration} 
  & Don’t have access & 32.2 (3.9) & 63 \\
  & Have access       & 33.2 (3.9) & 73 \\
  & \multicolumn{3}{r}{\textit{F (p) = 2.34 (0.129)}} \\
\hline
\end{tabular}
    \begin{tablenotes}
    \small
    \item SD = standard deviation; N = group size; F (p) = ANOVA F statistic (p-value). \\
    \textbf {p} < 0.05 is considered statistically significant, \textbf{p} < 0.1 is considered marginally significant.
    \end{tablenotes}
\end{table}

To provide additional context, we also examined participants’ reported usage of LLMs. Among those with access, usage frequency ranged from several times a day to occasional task-specific use. Most applied LLMs to programming tasks such as learning new concepts, writing or improving code, and debugging, while others reported using them for study planning, clarifying difficult topics, and exploring supplementary resources. 

\section{Discussion}
This study demonstrates the feasibility of adapting Snyder’s Hope Scale as a metric for aspirations in socio-politically unstable contexts. The adapted scale showed good reliability, and participants found it understandable and relevant, which supports its use in measuring educational aspirations. Our findings nuance the role of technology, particularly Large Language Models (LLMs), in shaping aspirations. While access to LLMs was not associated with overall aspiration levels, we observed a marginally significant association with the \textit{Avenues} subscale. This suggests that LLMs may function as a facilitator, expanding learners’ perception of possible routes to achieving their aspirations. Such an interpretation is consistent with prior work positioning technology as an amplifier of existing capacities in international development \cite{toyama2018needs, toyama2011technology}.

These insights align with broader scholarly calls to shift technology design from meeting basic needs toward actively supporting user aspirations \cite{toyama2018needs}. The adapted Hope Scale provides a metric for understanding user aspirations, enabling researchers and practitioners to evaluate whether technological interventions reinforce longer-term capacities to aspire, rather than merely addressing immediate educational challenges. Beyond educational contexts, aspiration-based assessments could inform technological interventions in employment and mental health. For instance, in employment support programs such as Harambee\footnotemark[1]\footnotetext[1]{\url{https://www.harambee.co.za/}}, aspiration assessments could guide personalized mentorship for those with limited perceived \textit{Avenues} or motivational reinforcement for those with lower \textit{Agency}. Similarly, in digital mental health, prior research has shown that mental states shape how individuals envision their futures \cite{pendse2019mental}; integrating the Hope Scale into such platforms could enable dynamic tracking of \textit{Agency} to provide timely, personalized encouragement.

At the same time, integrating LLMs into education raises important challenges. Recent studies have identified concerns around over-reliance on LLM-generated content, diminished critical thinking, the propagation of misinformation, and the reinforcement of biases embedded in models \cite{harvey2025don}. Considering these challenges is essential while utilizing LLMs to support learners’ aspirations equitably.

Finally, we acknowledge a limitation of this study. While the adapted Hope Scale demonstrated reliability, it primarily captures the cognitive dimensions of aspiration—\textit{Agency} and \textit{Avenues}—and may not fully capture the socio-environmental factors emphasized in development studies and HCI4D, such as community support and institutional barriers \cite{toyama2018needs, kumar2014facebook}. Future research should therefore complement this scale with methodological approaches that situate aspirations within broader social and political contexts.

\section{Conclusion}
This study shows that Snyder’s Hope Scale, when adapted to the programming context, is a feasible metric for understanding learners' aspirations among women in socio-politically unstable settings. While overall aspiration levels were not affected by LLM access, a marginal effect on the \textit{Avenues} subscale suggests that such technologies may expand learners’ perceived pathways to achieving their long-term goals. These findings underscore the importance of aspiration-based metrics in evaluating technological interventions and designing technologies that support not only immediate learning but also long-term educational and career aspirations. Future work should examine longitudinal effects, extend the adapted scale to other marginalized populations, and integrate qualitative methods to capture the broader socio-environmental dimensions of aspiration.

\section*{Acknowledgments}
IW and VC are supported by funding from the Alexander von Humboldt Foundation and its founder, the German Federal Ministry of Education and Research. 

\renewcommand{\refname}{References}
\renewcommand{\bibname}{References}

% {\let\clearpage\relax
% \let\cleardoublepage\relax
\appendix
\section{Appendix}

\begin{table}[H]
\centering
\setlength{\tabcolsep}{12pt}
\renewcommand{\arraystretch}{1.2}

\caption{Hope scale questions modified for programming. Items marked with † are fillers; as in the traditional scale, they are not scored and are excluded from analysis.}
\label{tab:hope-scale}

\begin{tabular}{@{}p{0.18\columnwidth} p{0.34\columnwidth} p{0.34\columnwidth}@{}}
\hline
\textbf{Subscale} & \textbf{Programming-Adapted Question} & \textbf{Original Hope Scale Question} \\
\hline

Hope scale --   Avenues & I can think of many ways to get out of debugging a programming problem. & I can think of many ways to get out of a jam. \\

Hope scale -- Agency & I energetically pursue my goals about programming. & I energetically pursue my goals. \\
\rowcolor{blue!10}
Hope scale$^{\dagger}$ & Working on a computer screen is tiring most of the time. & I feel tired most of the time. \\

Hope scale --   Avenues & There are lots of ways around any programming problem. & There are lots of ways around any problem. \\
\rowcolor{blue!10}
Hope scale$^{\dagger}$ & I give up quickly when I find a new idea. & I am easily drawn into an argument. \\

Hope scale --   Avenues & I can think of many ways to get the things that are most important to me in programming. & I can think of many ways to get the things in life that are most important to me. \\
\rowcolor{blue!10}
Hope scale$^{\dagger}$ & I worry about my poor posture while programming. & I worry about my health. \\

Hope scale --   Avenues & Even when others get discouraged, I know I can find a way to solve a programming problem. & Even when others get discouraged, I know I can find a way to solve the problem. \\

Hope scale -- Agency & My past programming experiences have prepared me well for my future. & My past experiences have prepared me well for my future. \\

Hope scale -- Agency & I've been pretty successful in life as a programmer. & I've been pretty successful in life. \\
\rowcolor{blue!10}
Hope scale$^{\dagger}$ & I usually find myself worrying about my programming assignments. & I usually find myself worrying about something. \\

Hope scale -- Agency & I meet the goals that I set for myself as a programmer. & I meet the goals that I set for myself. \\
\hline
\end{tabular}
\raggedright\footnotesize
\textbf{Response Options:} Responses were recorded on a five-point Likert scale — 1 = Strongly Disagree, 2 = Disagree, 3 = Neither Agree nor Disagree, 4 = Agree, 5 = Strongly Agree. \\
\end{table}

\end{document}